\documentclass[prb, twocolumn, showpacs, showkeys, superscriptaddress]{revtex4-1}

\usepackage{amssymb, amsfonts, amsmath}
\usepackage{setspace}
\usepackage{bbm}
\usepackage[english]{babel}
\usepackage{graphicx}
\usepackage[applemac]{inputenc}
\usepackage{color}
\usepackage{ulem}

\graphicspath{ {./Pictures/} }

\begin{document}

\newcommand{\bra}[1]{\ensuremath{\left\langle #1\r|}}
\newcommand{\ket}[1]{\ensuremath{\left|#1\r\rangle}}
\newcommand{\braket}[2]{\ensuremath{\left\langle #1\vphantom{#2}\r.\left|\vphantom{#1}#2\r\rangle}}
\newcommand{\proj}[1]{\ensuremath{\ket{#1}\bra{#1}}}

\newcommand{\mean}[1]{\ensuremath{\left\langle #1\r\rangle}}
\newcommand{\var}[1]{\ensuremath{\left\langle \left\langle #1\r\rangle \r\rangle}}

\newcommand{\cc}{^{\ast}}                						
\newcommand{\hc}{^{\dagger}}             					
\newcommand{\op}[1]{\hat{#1}}							
\newcommand{\ee}{\mathrm{e}}             					
\newcommand{\ii}{\mathrm{i}}             					
\renewcommand{\H}[0]{\hat{H}}  						
\renewcommand{\L}[0]{\mathcal{L}}  						

\newcommand{\comm}[2]{\left[ #1, #2 \right]} 				
\newcommand{\acomm}[2]{\left\{ #1, #2 \right\}} 			
\renewcommand{\Im}{\ensuremath{ \operatorname{Im} }}		
\renewcommand{\Re}{\ensuremath{ \operatorname{Re} }}		
\newcommand{\nn}{\nonumber}							
\newcommand{\abs}[1]{\ensuremath{ \left| #1 \right| }}		
\newcommand{\abss}[1]{\ensuremath{ \left| #1 \right|^{2} }}	
\newcommand{\diss}[1]{\mathcal{D}[ #1 ]}					
\renewcommand{\l}[0]{\left}
\renewcommand{\r}[0]{\right}

\def\vphi{\varphi}									
\newcommand{\dvpart}[2]{\frac{\partial #1}{\partial #2}}		
\newcommand{\ddvpart}[2]{\frac{\partial^2 #1}{\partial #2 ^2}}
\newcommand{\dvtot}[2]{\frac{d #1}{d #2}}
\newcommand{\PiPhi}[0]{\left( \frac{2\pi}{\Phi_{0}} \right)}

\newcommand{\eq}[1]{(\ref{#1})}

\newcommand{\red}{\color[rgb]{0.8, 0, 0}}
\newcommand{\orange}{\color[rgb]{1.0, 0.6, 0.0}}
\newcommand{\blue}{\color[rgb]{0.0, 0.0, 0.6}}

\newcommand{\ab}[1]{{\red AB:~#1}} 		
\newcommand{\jb}[1]{{\blue JB:~#1}} 		
\newcommand{\cm}[1]{{\orange CM:~#1}}		

\title{Detection and Manipulation of Majorana Fermions in Circuit QED}
	
\date{\today}

\author{Clemens M\"uller}
	\affiliation{D\'epartement de Physique, Universit\'e de Sherbrooke, Sherbrooke, Qu\'ebec, Canada, J1K 2R1}
\author{J\'er\^ome Bourassa}
	\affiliation{D\'epartement de Physique, Universit\'e de Sherbrooke, Sherbrooke, Qu\'ebec, Canada, J1K 2R1}
\affiliation{D\'epartement des Sciences de la Nature, C\'EGEP de Granby-Haute Yamaska, Granby, Qu\'ebec, Canada, J2G 9H7}
\author{Alexandre Blais}
	\affiliation{D\'epartement de Physique, Universit\'e de Sherbrooke, Sherbrooke, Qu\'ebec, Canada, J1K 2R1}

\begin{abstract}
	Motivated by recent experimental progress towards the measurement and manipulation of Majorana fermions with superconducting circuits,
	we propose a device interfacing Majorana fermions with circuit quantum electrodynamics.
	The proposed circuit	acts as a charge parity detector changing the resonance frequency of a superconducting $\lambda/4$-resonator 
	conditioned on the parity of charges on nearby gates.
	Operating at both charge and flux sweet spots, this device is highly insensitive to environmental noise. It 
	enables high-fidelity single-shot quantum non demolition readout of the state of a pair of Majorana fermions encoding a topologically protected qubit. 
	Additionally, the interaction permits the realization of an arbitrary phase gate on the topological qubit, closing the loop for computational completeness.   
	Away from the charge sweet spot, this device can be used as a highly sensitive charge detector 
	with a sensitivity better than $10^{-4} \text{e} / \sqrt{\text{Hz}}$ and bandwidth larger than $1$~MHz.
\end{abstract}


\maketitle

\section{Introduction}

	Pairs of Majorana fermions have been put forward as candidates for topologically protected quantum computation~\cite{Kitaev:2007a, Nayak:2008a}, and
	have attracted much attention from both theoretical and experimental groups~\cite{Sau:2010a, Oreg:2010a, Alicea:2011a, Mourik:2012a, Das:2012a, Rokhinson:2012a, Churchill:2013a}.
	First experimental indications were recently reported that the search for these fermions has been successful 
	in semiconductor wires on superconducting surfaces~\cite{Mourik:2012a, Das:2012a, Rokhinson:2012a, Churchill:2013a}. 
	In these systems, selective gating of a strongly spin-orbit coupled 1D semiconductor on top of a superconducting substrate and under the influence of a magnetic field
	allows the nucleation and displacement of pairs of Majorana modes at the endpoints of topologically nontrivial regions~\cite{Sau:2010a, Oreg:2010a}. 
	Since individual Majorana fermions are Ising anyons, pairs of these fermions have been proposed as topologically protected qubits, 
	for which the majority of single- and two-qubit operations can be performed via braiding~\cite{Alicea:2011a}.
	Importantly, the nonlocal fermion defined by two Majorana endmodes still carries the charge of the underlying carriers, i.e., one electron charge. 
	
	On the other hand, superconductor based technology for use in quantum computation 
	has had tremendous success in recent years~\cite{Neeley:2010a, DiCarlo:2010a, steffen:2013a, Devoret:2013a}.
	One well studied architecture is circuit quantum electrodynamics (cQED)~\cite{Blais:2004a, Wallraff:2004a}, 
	in which superconducting qubits interact strongly with the electric or magnetic fields of a
	superconducting resonator. In this area, much of the recent progress is due to the development of high-fidelity, 
	quantum nondemolition (QND) qubit readout schemes~\cite{Mallet:2009a, Johnson:2012a, Riste:2012a, steffen:2013a, Riste:2013a}, 
	based on measurements of the qubit-state dependent resonator frequencies~\cite{Blais:2004a}.
	Since the nucleation of Majorana fermions occurs on a standard BCS superconductor, it is natural to bridge the gap between these topological excitations and cQED.
	
	Previous authors have proposed to interface Majorana fermions in semiconductor wires and cQED by coupling the Majorana fermions to transmons~\cite{Hassler:2011a, Hyart:2013a} 
	or charge tuneable flux qubits~\cite{Hassler:2010a, Bonderson:2011a} that are themselves coupled via Jaynes-Cummings interaction to superconducting resonators.
	Additional proposals include coupling semiconductor wires to a superconducting cavity in order 
	to induce a photon-mediated effective interaction between Majorana fermions~\cite{Schmidt:2012a}
	or to generate squeezing of the resonator field~\cite{Cottet:2013a}. 
	It was also proposed to utilize the $4\pi$-periodic Josephson effect in conjunction with a fluxonium circuit for Majorana qubit detection~\cite{Pekker:2013a}.
	In many of these proposals~\cite{Hassler:2010a, Hyart:2013a, Hassler:2011a, Bonderson:2011a, Pekker:2013a}, 
	the decoherence properties of the underlying qubit used for the interaction to microwave photons are crucial when trying to achieve fast readout. 
	This is the case for example for the top-transmon~\cite{Hassler:2011a}, which during readout is operated far from the noise-insensitive transmon regime~\cite{Koch:2007a}. 
	
	Here, we suggest a device for measurement and manipulation of Majorana fermion qubits that is only weakly affected by decoherence
	while still allowing fast QND readout. 
	In our design, the superconducting circuitry acts as a purely passive element, with no internal dynamics which might be influenced by relaxation or dephasing.
	Additionally, the device is operating at sweet spots with respect to both charge as well as flux, and is thus to first order insensitive to fluctuations in these external parameters. 
	Readout of the charge parity is achieved by a standard measurement of the resonance frequency of a superconducting transmission line resonator, 
	which is here conditioned on the state of a nearby Majorana fermion qubit. 

	The paper is structured as follows. Section~\ref{sec:Idea} introduces the principle idea of our proposal 
	while Sec.~\ref{sec:JCPM} details how this can be implemented with superconducting circuits and presents the main result of this paper, 
	namely the tuneability of the transmission line mode frequency with applied charge. 
	Section~\ref{sec:Noise} is devoted to exploring the sensitivity of our design to external perturbations and fabrication imperfections. 
	We end the paper with a short discussion and summary of the results.
	Appendix details the derivation of the device Hamiltonian as well as discusses the parameters necessary for experimental realization. 
	There we also give details on the working of the Aharonov-Casher effect in our proposal as well as discuss how the same device can be used as a dynamical charge detector.

\section{Charge-tuneable inductance\label{sec:Idea}}
	
	\begin{figure*}[tbp]
		\centering
		\includegraphics[width=.9\textwidth]{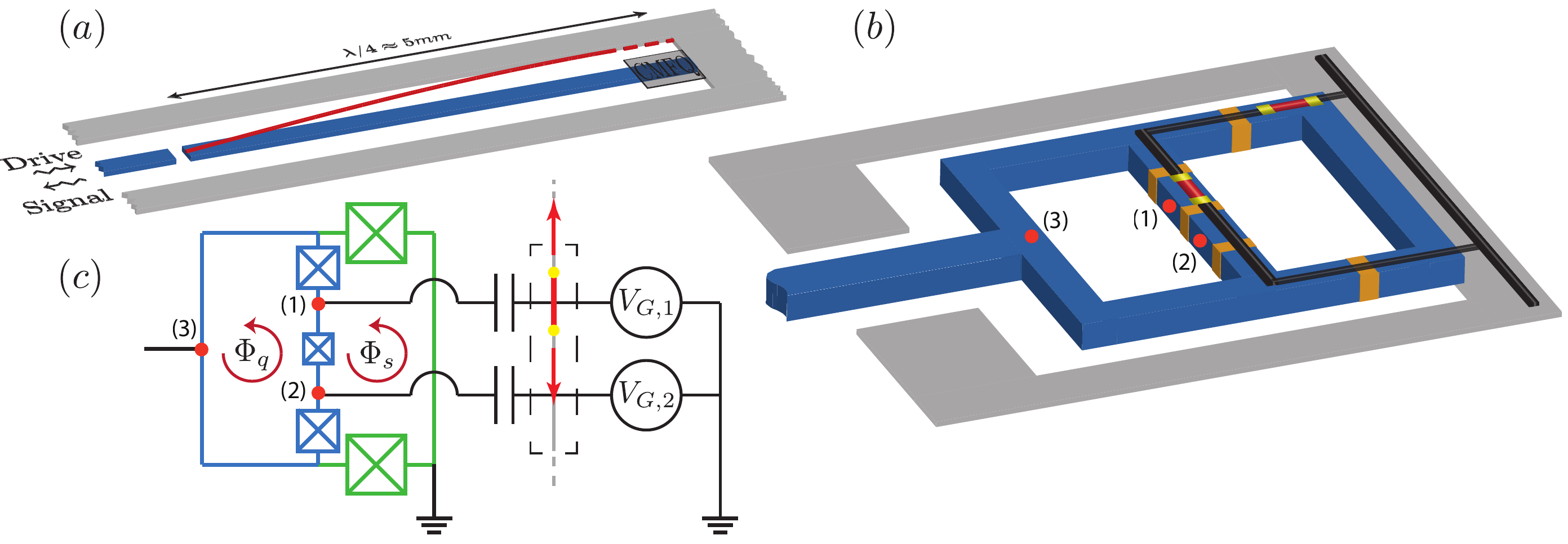}
		\caption{(Color online) Schematic illustration of the JCPM. 
			\textbf{(a)} A superconducting $\lambda/4$ resonator is terminated to ground via a 
			tuneable inductance, in our proposal realized by the CMFQ, indicated here as a gray square. 
			Measurement is realized via driving through the capacitive port on the left 
			and monitoring the reflected signal. The red lines indicate the current distribution for the first fundamental mode in the structure.
			\textbf{(b)} Sketch of the CMFQ including a semiconductor wire network supporting Majorana fermions.  
			The device itself consists of a flux qubit symmetrically coupled to the two arms of a SQUID. 
			Blue indicates superconducting wires interrupted by Josephson junctions in orange, the grey area shows the surrounding superconducting groundplanes.
			A semiconductor wire network is indicated as black lines with red parts illustrating topologically nontrivial regions terminated by Majorana endmodes in yellow.
			\textbf{(c)} Circuit diagram of the CMFQ including the interaction region in black modelled as a set of capacitors connected to voltage sources / charges. 
			Boxes indicate Josephson junctions and the different colors denote the SQUID (green) and flux qubit (blue) parts of the circuit.	 
			For ease of identification, the numbered red dots in (b) and (c) indicate equivalent points in the circuit. 
		}
		\label{fig:JPMFull}
	\end{figure*}

	Our proposal is related to flux-tuneable microwave cavities commonly used in cQED architectures~\cite{Wallquist:2006a}, c.f. Fig.~\ref{fig:JPMFull}~(a). 
	There, a $\lambda/4$-resonator is terminated to ground via a superconducting quantum interference device (SQUID) loop that plays the role of a flux-tuneable inductance. 
	This change of inductance modifies the electrical length of the resonator and in turn its resonance frequency~\cite{Wallquist:2006a,Sandberg:2008a}.
	This can be simply pictured as a standard LC-resonator with an additional tuneable inductance. 
	The oscillator's resonance frequency is then $\omega_{r} = 1/ \sqrt{(L + L_{J}) C}$ 
	where $L_{J}$ is the tuneable inductance.
	Changing $L_{J}$ leads to a change $\delta\omega_{r}$ in the resonator frequency $\omega_{r}$ according to
	\begin{align}
		\frac{\delta\omega_{r}}{\omega_{r}} = - \frac12 \frac{\delta L_{J}}{L + L_{J}} = -\frac{p_{L}}{2} \frac{\delta L_{J}}{L_{J}}, 
	\end{align}
	where $\delta L_{J}$ is the change in the tuneable inductance of the circuit 
	and $p_{L} = L_{J} / (L + L_{J})$ is the inductive participation ratio. 
	In the circuit of Fig.~\ref{fig:JPMFull}~(a), this tuneable inductance can be realized by a SQUID loop.  
	In that case, the Josephson inductance $L_{J} \propto 1/E_{J}$, with $E_J$ the Josephson energy, depends on the magnetic flux threading the loop~\cite{Makhlin:1998a}.
	
	In contrast to flux tuneable devices we aim here at designing a charge tuneable inductance. 
	To this end we make use of the Aharonov-Casher (AC) effect, the charge-flux dual of the Aharonov-Bohm effect
	in a superconducting flux qubit~\cite{Friedman:2002a, Chirolli:2006a}. 
	As discussed in more details in \ref{sec:AharonovCasher}, the AC effect can be important in flux qubits because of the strong dependence of the tunneling amplitude 
	between different wells of the qubit potential energy. 
	In the charge tuneable regime~\cite{Chirolli:2006a}, there exist two competing tunneling paths each of which will acquire a different phase 
	conditioned on the charges present on the qubit islands, i.e. the small superconducting regions between the qubits Josephson junctions. 
	The AC effect is periodic in the two island charges with a period of $2$e and the maximal phase difference is achieved for a single applied charge to either one of the islands.
	Here, e is the charge of a single electron.
	This allows to tune the flux qubit's transitions frequency with applied charges or gate voltages~\cite{Chirolli:2006a, Tiwari:2007a, Hassler:2010a}. 
	Because of the $2$e periodicity, it can serve as a natural detector of the charge parity on the qubit islands.
	
	To take advantage of this effect, we propose to modify the standard flux-tuneable resonator by including a charge-sensitive flux qubit in the terminating SQUID 
	of a flux-tuneable $\lambda/4$-resonator.
	Our circuit provides a highly symmetric coupling of the resonator to the qubit loop, 
	and leads to vanishing cross-coupling between excitations in the resonator and in the qubit. 
	On the other hand, as we will show below, it results in a strong tuneability of the oscillator resonance frequency as a function of the gate charge applied to the qubit islands.
	The proposal can be broken down into two parts, cf. Fig.~\ref{fig:JPMFull}. 
	The first is a measurement circuit in which the level splitting of a flux qubit is tuned by an applied charge, 
	something which we will refer to as a charge modulated flux 1ubit (CMFQ) for simplicity. 
	Details of the CMFQ as well as its circuit diagram are depicted in Figs.~\ref{fig:JPMFull}~(b) and~\ref{fig:JPMFull}~(c).
	By terminating a superconducting $\lambda/4$-resonator to ground via the CMFQ, we arrive at the 
	Josephson charge parity meter (JCPM), illustrated in Fig.~\ref{fig:JPMFull}~(a). 
	
\section{Working principle of the JCPM\label{sec:JCPM}}
	
	\begin{figure}[tbp]
		\centering
		\includegraphics[width=0.9\columnwidth]{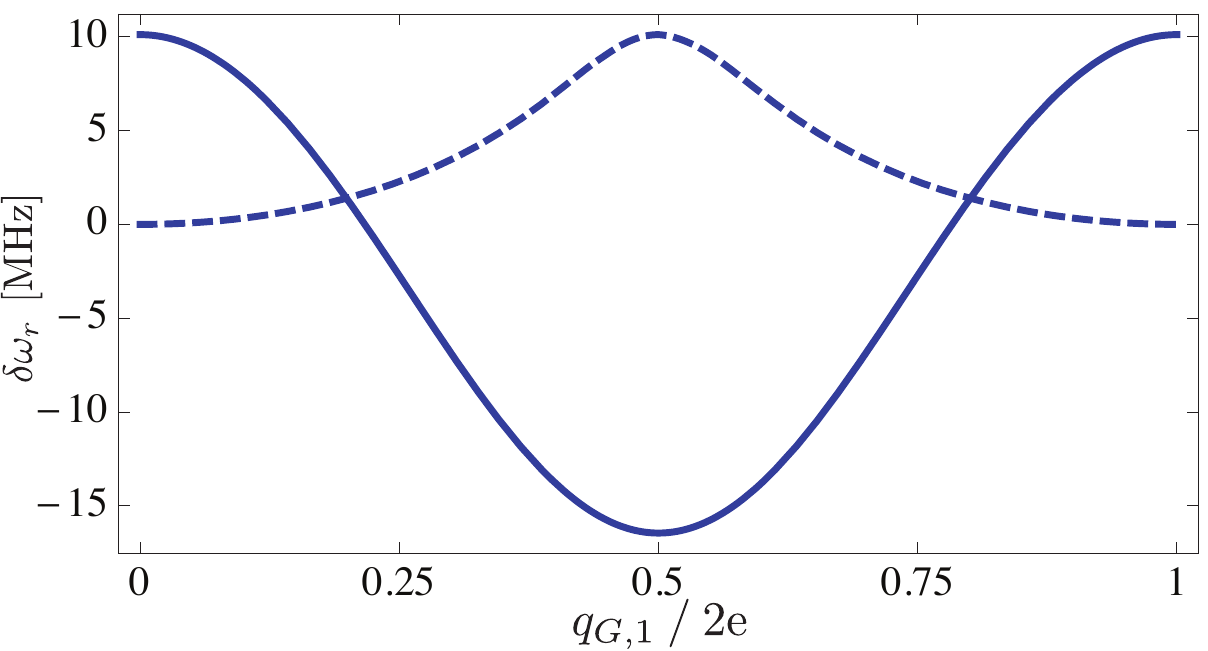}
		\caption{(Color online) Modulation of the resonator frequency pull $\delta\omega_{r}$ as a function of charge $q_{G,1}$ on the lower qubit island.
			The magnetic field is fixed such that the device is at its flux sweet spot with $\Phi_{q}  = \Phi_{s} = \Phi_{0}/2$.
			The solid line is for a single charge on the upper qubit island, $q_{G,2} = \text{e}$, the dashed line for $q_{G,2} = 0$. 
			The parameters are presented in \ref{sec:Parameters} and are well within the standard toolbox of circuit QED.
		}
		\label{fig:ResFreq}
	\end{figure}
	
	As shown schematically in Fig.~\ref{fig:JPMFull}~(b), the circuit
	consists of two superconducting loops, one a SQUID loop with two Josephson junctions, the other a flux qubit with three junctions. These are inserted between the endpoint of a
	$\lambda/4$ transmission line resonator (indicated by the resonator center pin arriving from the left) and the surrounding superconducting groundplanes (grey area). 
	A network of semiconductor wires supporting Majorana endmodes is indicated on top of the superconducting structure by the black lines.
	Topologically nontrivial regions on the wire, represented by red areas and terminated by Majorana endmodes in yellow, can be moved 
	along the network via depletion gates (not shown)~\cite{Alicea:2011a, Mourik:2012a}. 
	The charge sensitive regions of the circuit are the two islands of the flux qubit, indicated as (1) and (2) in Figs.~\ref{fig:JPMFull}~(b) and~\ref{fig:JPMFull}~(c). 
	They are the only parts of the circuit with an nonnegligible electrostatic energy, 
	and are capacitively coupled to the nearby semiconductor network to serve as charge sensors. 
	This is indicated by the coupling capacitors in Fig.~\ref{fig:JPMFull}~(c).
	A pair of Majorana fermions close to one of the qubit islands will then induce a charge on the islands 
	depending on whether its parent Dirac fermion mode is occupied or not~\cite{Hassler:2010a, Hassler:2011a}. 
	The two islands are here completely equivalent. In the following we will use the upper one, marked as (1), for the measurement of the charge.
	The lower island, marked as (2), might be biased with a static voltage to improve the readout contrast, as illustrated by the voltage source in Fig.~\ref{fig:JPMFull}~(c).
	
	In the absence of tunneling, the states of a flux qubit near its flux sweet spot, $\Phi_{q} = \Phi_{0}/2$, are characterized by a super-current flowing 
	clockwise or counter-clockwise in the loop~\cite{Mooij:1999a}. The circuit of Fig.~\ref{fig:JPMFull}~(b)
	couples both current states of the qubit loop symmetrically to currents flowing from the resonator to ground via the CMFQ.
	As a result, Jaynes-Cummings interaction between excitations in the resonator and in the qubit is fully suppressed, 
	since this interaction is mediated by the qubit magnetic dipole moment. 
	
	Using the approach of Ref.~\onlinecite{Bourassa:2012a}, we derive the full Hamiltonian of the JCPM in~\ref{sec:Lagrangian}.
	In particular, the coupling between resonator and CMFQ degrees of freedom is described by
	\begin{align}
		\H_{\text{Q-Res}} &= 2 E_{J,q} \sin\psi \: \sin{\frac{\vphi_{+}}{2}}\cos{\frac{\vphi_{-}}{2}}  \nn\\
			&+ 2 E_{J,q} \l( \cos\psi - 1 \r) \cos{\frac{\vphi_{+}}{2}}\cos{\frac{\vphi_{-}}{2}}  \,,
		\label{eq:LInd}
	\end{align}
	where $\psi$ characterizes the resonator field at the input of the CMFQ, 
	and the phases $\vphi_{+}$ and $\vphi_{-}$ describe the dynamics of the flux qubit circuit. 
	$E_{J,q}$ is the Josephson energy of the two identical outer qubit junctions.
	In the limit of infinite Josephson energy of the SQUID junctions, $E_{J,s} \rightarrow \infty$, 
	the CMFQ would act as a simple short to ground and thus we would find $\psi = 0$.
	For our purposes we choose large SQUID junctions $E_{J,s}$, such that $\psi \ll 1$, while still maintaining a significant inductive participation ratio $p_{L}$.
	In this limit, we expand Eq.~\eq{eq:LInd} to obtain
	\begin{align}
		\H_{\text{Q-Res}} &\approx 2 E_{J,q} \l\{ \psi \: \sin{\frac{\vphi_{+}}{2}}\cos{\frac{\vphi_{-}}{2}} + \psi^{2} \: \cos{\frac{\vphi_{+}}{2}}\cos{\frac{\vphi_{-}}{2}}\r\}  \,.
		\label{eq:LInd2}
	\end{align}
	The first term of this expression leads to coupling between flux qubit excitations and resonator photons, while the second term renormalizes the resonator frequency. 
	This can be seen more clearly by expressing $\psi$ in terms of creation $(a)$ and annihilation $(a^\dag)$ operators of resonator photons, c.f.~\ref{sec:Lagrangian}, to find
	\begin{equation}\label{eq:HqresOptics}
		\begin{split}
		\H_{\text{Q-Res}} & \approx \l( a\hc + a  \r) \sum_{i} g_{i} \l( \sigma_{+}^{(i)} + \text{h.c.} \r) \\
						&\quad+ a\hc a\sum_{i} \delta\omega_{r,i} \ket i \bra i  \,,
		\end{split}
	\end{equation}
	with $\ket i$ the $i$-th eigenstates of the flux qubit and $\sigma_{+}^{(i)} = \ket{i+1} \bra{i} $. 
	We have also defined the Jaynes-Cummings coupling strength $g_{i} \propto \bra{i} \sin{\frac{\vphi_{+}}{2}}\cos{\frac{\vphi_{-}}{2}} \ket{i+1} + \text{h.c.}$
	and the resonator frequency shifts $\delta\omega_{r,i} \propto \bra i \cos{\frac{\vphi_{+}}{2}}\cos{\frac{\vphi_{-}}{2}} \ket i $.
	Both depend on the flux qubit eigenstates and are therefore sensitive to charges on the flux qubit islands.
	In practice, the coupling $g_{i}$ is relevant only if the resonator and flux qubit frequencies are close to resonant, $\Delta = \omega_{q} - \omega_{r} \lesssim g$, 
	where $\omega_{q}$ is the qubit level splitting and $\omega_{r}$ the resonator frequency.
	As we will show in the following Sec.~\ref{sec:Noise}, for our circuit design and parameters we find $g / \Delta < 10^{-2}$ at all operating points
	such that no excitations are exchanged between resonator and flux qubit. 
	Additionally, thermal excitations of the flux qubit can be neglected since the qubit transition frequency can easily be chosen such that $\hbar \omega_{q} \gg k_{B} T$.
	As a result, and as will be discussed in more details below, the flux qubit remains at all time in its ground state 
	and the main effect of the CMFQ on the resonator is to change the resonator frequency by $\delta\omega_{r} \equiv \delta\omega_{r,0}$. 
	Importantly, given the dependence of the qubit's eigenstates on charge, the resonator frequency is then modified by the presence or absence of a charge on the flux qubit island. 
	This is the effect that we propose to take advantage of.
	
	Figure~\ref{fig:ResFreq} shows this resonator frequency pull $\delta\omega_{r}$ as a function of gate charge $q_{G,1}$ on the upper qubit island.
	The parameters used here are similar to those of many recent experiments and are presented in~\ref{sec:Parameters}. 
	In particular, we choose a flux bias point of $\Phi_{s} = \Phi_{q} = \Phi_{0} / 2$ such that both the SQUID and the flux qubit are at their respective flux sweet spots. 
	In this situation, they are both first-order insensitive to magnetic flux variations.
	Incidentally this means that the area of the two corresponding loops should ideally be chosen to be of equal size.
	The solid line in Fig.~\ref{fig:ResFreq} corresponds to the case where the CMFQ is biased with an additional voltage on the lower island 
	equivalent to a single charge on that island, $q_{G,2} = $~e. 
	This choice increases the frequency pull but in practice is not necessary. 
	Indeed, as is made evident by the dashed line, without this biasing the frequency pull is reduced but still large enough to be easily detected.
	With biasing, we find a frequency difference of the JCPM between different charge parity states on the first island of $\delta\omega_{r} > 25$~MHz.
	This frequency change is well above the typical qubit-cavity pull and photon relaxation rate $\kappa$ in cQED 
	making fast and high-fidelity readout possible~\cite{Johnson:2012a, Riste:2012a}. 
	
	Apart from allowing charge sensitivity, the inclusion of the nonlinear Josephson elements in the resonator 
	introduces a small Kerr-type nonlinearity $K$ to the resonator modes~\cite{Bourassa:2012a}. This corresponds to adding a term $\sim K \l( a\hc a \r)^{2}$ to the resonator Hamiltonian.
	As is described in~\ref{sec:Parameters}, the magnitude of $K$ will also be modulated with the applied gate-charge
	and it is of the right order of magnitude to be exploited in bifurcation readout~\cite{Mallet:2009a, Ong:2011a}. 

\section{Robustness to noise and fabrication imperfections \label{sec:Noise}}

	As mentioned above, the design of the JCPM circuit ensures that its degrees of freedom remain in their ground state, such that it will act as a purely passive detector, 
	with no internal dynamics that might be susceptible to decoherence. 
	In this section, we provide more quantitative arguments supporting this claim, and also discuss tolerance to noise and deviation from optimal parameters. 
		
	\begin{figure}[tbp]
		\begin{center}
			\includegraphics[width=.9\columnwidth]{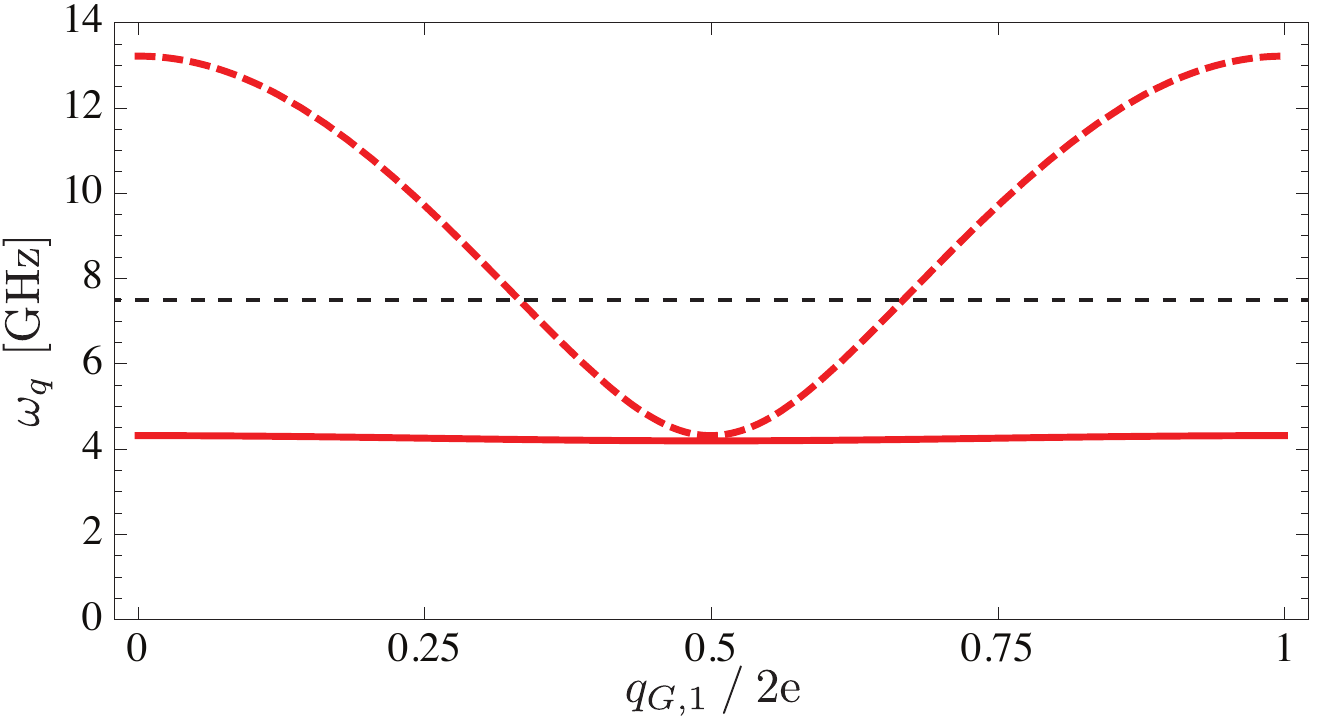}
			\caption{(Color online) Level splitting $\omega_{q}$ of the two lowest levels of the qubit as a function of island charge $q_{G,1}$ at 
				the flux sweet spot, $\Phi_{q} = \Phi_{s} = \Phi_{0}/2$, and for two different charge bias points, $q_{G,2} = e$ (solid) and $q_{G,2} = 0$ (dashed).
				The qubit energy is always well above typical experimental temperatures of $\sim 20$~mK, corresponding to $\sim 400$~MHz,
				and the probability of thermal excitation is negligible. The bare resonator frequency is chosen to be 7.5~GHz as indicated by the dotted line.
			}
			\label{fig:QubitFreq}
		\end{center}
	\end{figure}

	Fig.~\ref{fig:QubitFreq} shows the level splitting $\omega_{q}$ of the two lowest qubit levels as a function of the gate charge $q_{G,1}$ for the parameters given in~\ref{sec:Parameters}. 	
	The qubit splitting is well above the thermal floor of typical cQED experiments, about $\sim20$~mK corresponding to $\sim400$~MHz, over the whole range.
	Thus thermal excitation of the flux qubit excited state can be safely neglected for these parameters. 
	Another possible source of excitation for the flux qubit is exchange of energy with the resonator caused by the first term of Eq.~\eqref{eq:HqresOptics}. 
	The effect of this term is however nonperturbative only when the qubit-resonator detuning $\Delta = \omega_{q} - \omega_{r}$ is smaller than the coupling $g$. 
	As illustrated in Fig.~\ref{fig:QubitFreq}, for the charge bias point $q_{G,2} = 0$, the qubit transition frequency crosses the resonator frequency $\sim 7.5$~GHz 
	(black dotted line) at two charge bias points. In practice, the resonator will be kept in its ground state during all Majorana fermion manipulations 
	(corresponding to charge rearrangement on the qubit islands) and will only be populated at the time of the readout. 
	As a result, these crossings will not cause excitations of the flux qubit which will stay in its ground state.
		
	\begin{figure}[tbp]
		\centering
		\includegraphics[width=0.9\columnwidth]{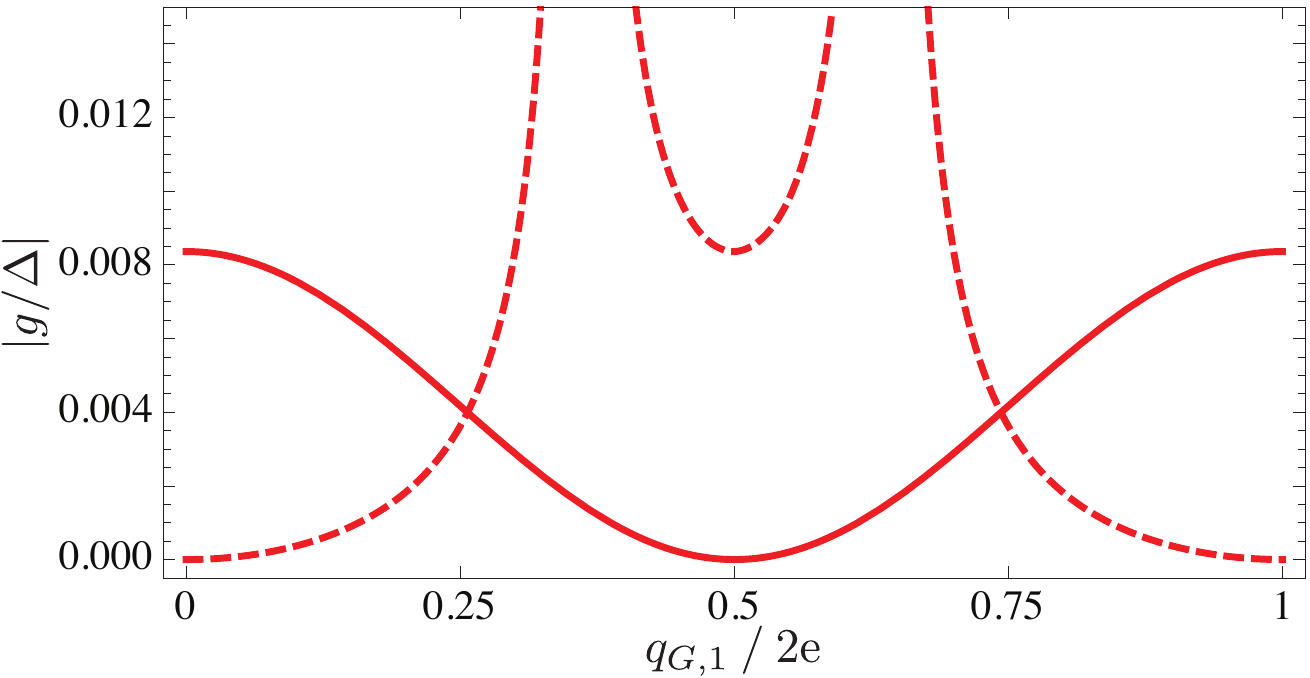}
		\caption{(Color online) Ratio of coupling $g$ to qubit-resonator detuning $\Delta = \omega_{q} - \omega_{r}$ 
			as a function of gate charge $q_{G,1}$ at the flux sweet spot of the device, $\Phi_{q}  = \Phi_{s} = \Phi_{0}/2$. 
			The solid line corresponds to $q_{G,2} =$~e (solid) while the dashed line to $q_{G,2} = 0$.
			At the two operating points of the JCPM, $q_{G,1} = 0$ and $q_{G,1} =$~e, we find $g / \Delta< 10^{-2}$ in both cases, 
			demonstrating that the symmetric circuit design leads to vanishing Jaynes-Cummings coupling between qubit and resonator excitations.
		}
		\label{fig:ResgDelta}
	\end{figure}
	
	Fig.~\ref{fig:ResgDelta} presents the ratio $|g/\Delta|$ as a function of $q_{G,1}$. Apart from the two divergences expected from the above discussion, 
	we find that the effect of the coupling $g$ to be perturbative. In particular, at the operating points for charge parity detection, where $q_{G,1}$ is either even or odd, 
	we find $g / \Delta < 10^{-2}$ for both $q_{G,2} = 0$ (dashed line) and $q_{G,2} = e$ (solid line).
	Interestingly, the resonance condition which is not ideal for charge detection could be used to facilitate entanglement between resonator photons and topological Majorana fermion qubits.

	\begin{figure}[tbp]
		\centering
		\includegraphics[width=0.9\columnwidth]{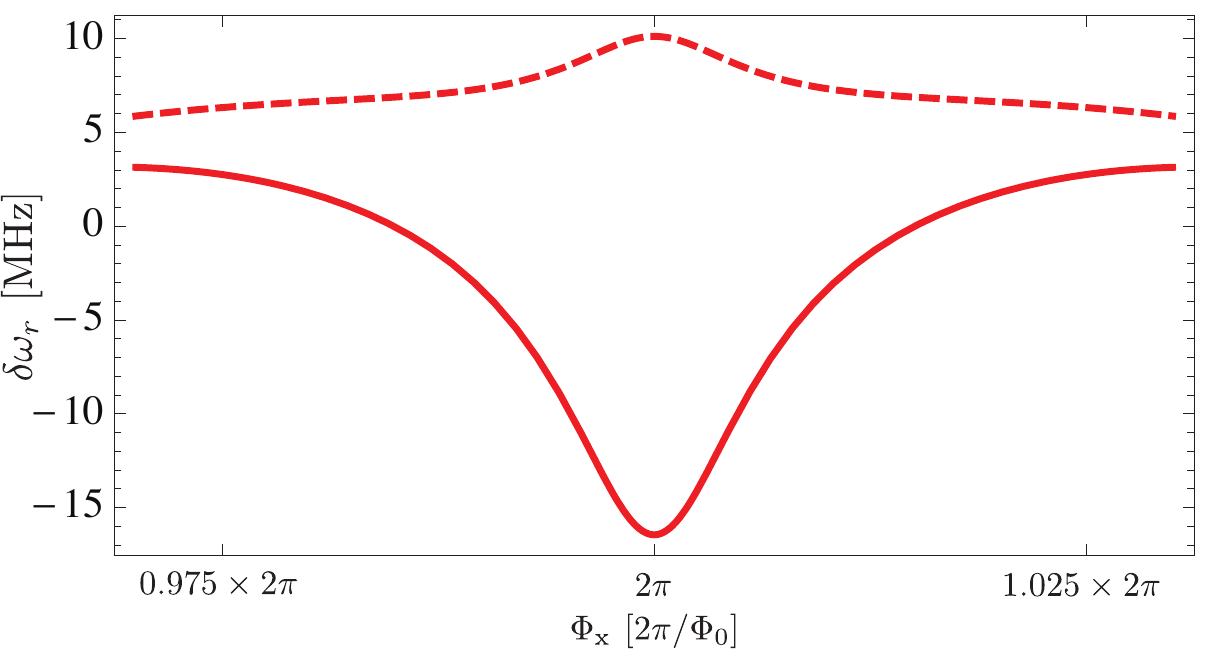}
		\caption{(Color online) Change in the JCPM resonance frequency $\delta\omega_{r}$ 
			as a function of the total external magnetic flux $\Phi_{\mathrm{x}}$. 
			We show the modulation of $\omega_{r}$ with respect to the state with zero charges on both qubit islands. 
			The solid line is for one of the JCPM working points, with $q_{G,1} = 0$ and $q_{G,2} =$~e, while the dashed line is for the other working point, 
			$q_{G,1} = q_{G,2} = $~e. At flux bias $\Phi_{x} = \Phi_{0}$, we find a sweet spot in both cases, where the resonance frequency $\omega_{r}$
			to first order does not depend on flux anymore.
		}
		\label{fig:DoubleSweet}
	\end{figure}
	
	\begin{figure}[tbp]
		\centering
		\includegraphics[width=0.9\columnwidth]{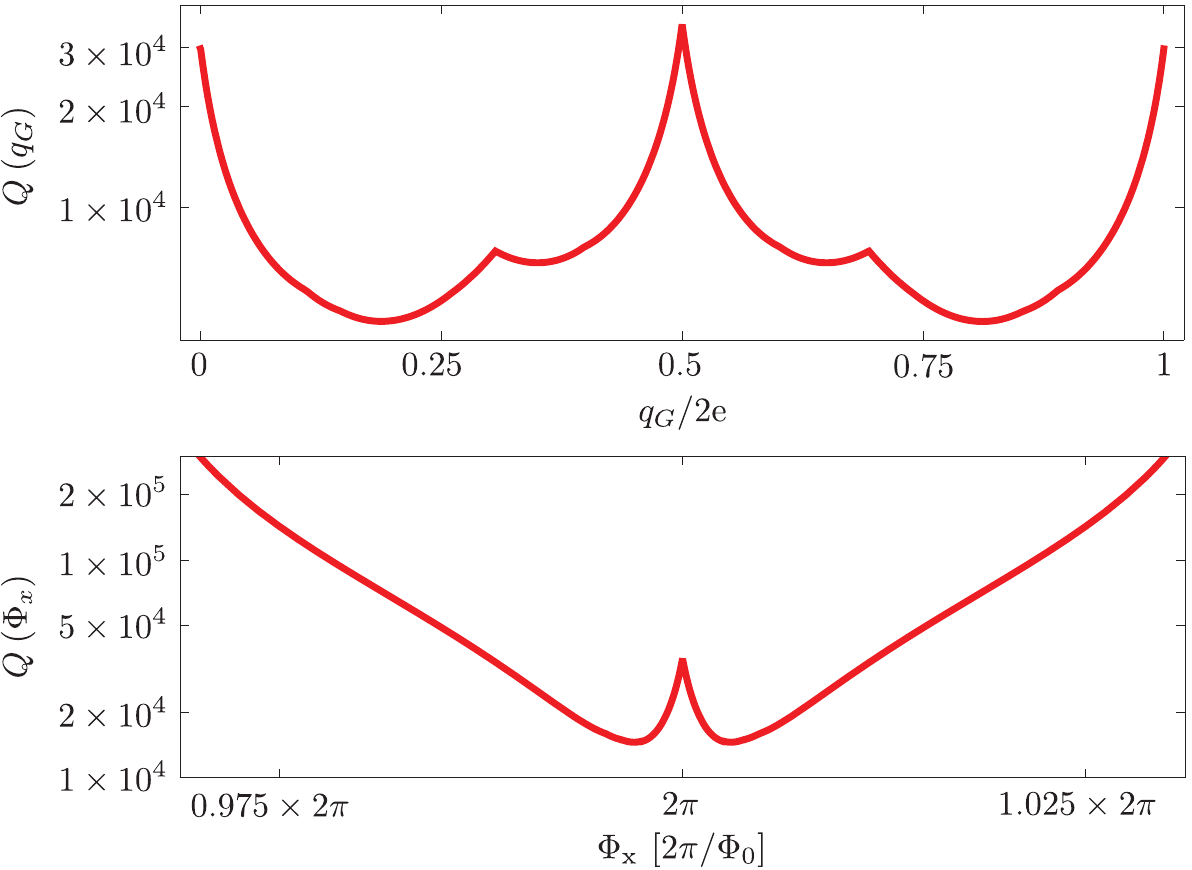}
		\caption{(Color online) Logarithmic plot of the resonator quality factor $Q$ as defined in the text with charge noise amplitudes of $\delta q = 10^{-2}$~e and 
			flux noise amplitude of $\delta\Phi_{\mathrm{x}} = 10^{-4} \Phi_{0}$. We show the quality factor as function of gate charge (top - equivalent for both islands) 
			and as function of total flux $\Phi_{\text{x}}$ through the CMFQ (bottom). 
		}
		\label{fig:QJPM}
	\end{figure}
	
	We now focus on the sensitivity to fluctuations in the external bias parameters, which we want to minimize. 
	These parameters are the two gate charges $q_{G,1/2}$ and the two fluxes $\Phi_{q/s}$.
	For parity measurement, the CMFQ is biased at $\Phi_{q} = \Phi_{s} = \Phi_{0} / 2$ and $q_{G_{2}} =$~e, 
	while the charge $q_{G,1}$ on the upper qubit island is either an even or odd number of electron charges.
	From Fig.~\ref{fig:ResFreq}, we see that the frequency pull as a function of charge on the qubit islands is constant around the operating points defined above.
	In Fig.~\ref{fig:DoubleSweet} we show the resonator frequency pull $\delta\omega_{r}$ as a function of the total external flux $\Phi_{x} = \Phi_{q} + \Phi_{s}$.
	These figures illustrate that the chosen bias conditions for external flux as well as gate charges correspond to sweet spots in the frequency dependence, 
	rendering the device to first order insensitive to noise in both control parameters. 
	
	To better quantify the resistance to noise, we define a quality factor of the JCPM as 
		\begin{align}
		Q = \omega_{r} \l( \sum_{x} \dvpart{\omega_{r}}{x} \delta x + \sum_{\{ x,y \}} \frac{\partial^{2} \omega_{r}}{\partial x \partial y} \delta x \delta y \r)^{-1} \,,
	\end{align}
	where $x$ and $y$ denote the different noise sources, i.e., charge noise on $q_{G,1/2}$ and flux noise on $\Phi_{x}$ with the noise amplitudes $\delta x$ and $\delta y$.
	This quality factor defines the robustness of the resonator frequency against noise in any of the external parameters 
	and thus is a figure of merit for the stability of the device operation.
	As illustrated in Fig.~\ref{fig:QJPM}, for a conservative choice of noise amplitudes of $10^{-2}$~e in charge~\cite{Vion:2002a} 
	and $10^{-4} \Phi_{0}$ in flux~\cite{Koch:2007a} the quality factor is larger than $10^{4}$ 
	at all operating points for charge parity detection and never falls below $10^{3}$ for all values of the input parameters.
	
	It is also useful to define a signal-to-noise ratio (SNR) as the quotient of the frequency shift (the signal) 
	over the induced frequency noise due to fluctuations in the bias parameters as defined above.  
	Defined this way, the SNR takes the form 
	\begin{align}
		\text{SNR} = \frac{\abs{\omega_{r}(0, \text{e}) - \omega_{r}(\text{e}, \text{e})}}{\abs{\text{max}(\delta\omega_{r})}} \approx 5 \times 10^{2}
	\end{align}
	where $\omega_{r}(q_{G,1}, q_{G,2})$ is the JCPM resonance frequency conditioned on the charges $q_{G,1/2}$ on the flux qubit islands
	and $\text{max}(\delta\omega_{r})$ is the maximum variation of $\omega_{r}$ due to noise with the same amplitudes as defined before. 
	With this large value of the SNR, we expect the frequency change to be readily measurable using standard microwave measurement techniques. 
	
	Our scheme depends strongly on the fact that the symmetric circuit design does not lead to any coupling between qubit and resonator excitations.
	The vanishing direct coupling is conditioned on the equality of the junctions parameters, 
	most importantly the outer two qubit junctions. To confirm that operation of the JCPM is not too sensitive on fabrication imperfections, we performed simulations
	where the parameters of the qubits junctions, $E_{J,q}$ and $C_{q}$ 
	varied randomly inside a gaussian distribution with a standard deviation of $5\%$, 
	a value which can be achieved for junction fabrication on the same chip~\cite{Fink:2009a}.
	Out of this ensemble of devices, more than $60\%$ showed parameters suitable for use as charge parity detector, specifically large qubit splitting $\omega_{q} >  2$~GHz 
	and a relative variation in Josephson energy and thus inductance $\delta E_{J} / E_{J} > 5\%$. 
	Relaxing the constraints on homogeneity of the junctions to a standard deviation of $10\%$, we still find an overall yield larger than $35\%$.
	
	In the proposal of Ref.~\onlinecite{Alicea:2011a}, a Majorana fermion network for quantum computation is realized by a 2D grid of semiconductor wires 
	which are statically biased by a collection of nearby voltage gates.  
	The presence of additional electrostatic gates close to the CMFQ circuit has the potential to disturb its operation by inducing unwanted charges. 
	However, their only effect on the CMFQ will be to provide a set of constant charge offsets during the measurement of the resonator frequency.
	In this situation the wire network will be biased such that a single pair of Majorana fermions resides on top of one of the qubit islands, 
	with the rest of the network in the topologically trivial state. 
	In all other computational situation, when topologically nontrivial regions are moved along the wires, the JCPM will be inactive 
	and therefore insensitive to the effect of the gates.
	When setting up the device, the effect of such a network of gates can then be calibrated for.
	
	Finally, much interest was recently devoted to understanding the effect of quasiparticle induced relaxation and dephasing in superconducting qubits. 
	It was found that the temperature dependence of the relaxation rate of transmon qubits is readily explained when considering the effect of interactions 
	with quasiparticles tunneling across the qubits Josephson junctions~\cite{Catelani:2011a}. 
	In our case, since the circuit will be resting in its ground state at all times, no energy is available to be absorbed by the bath of quasiparticles. 
	Additionally quasiparticle tunneling might lead to random frequency shifts of the circuit energy levels, which in the case of qubits lead to dephasing~\cite{Catelani:2012a}.
	This effect is small in charge insensitive devices like our proposed circuit, which are operated at sweet spots 
	where it energy levels to first order do not depend on the induced charge~\cite{Koch:2007a}. 
	However, if the charge noise present on either one of the qubit islands reaches a significant fraction of an electron charge 
	the frequency shift will be significant, and the operation will be disrupted. 
	This is a natural limitation of our proposal stemming from the fact that the circuit is a charge parity meter.
	
\section{Discussion and Summary}

	Similar to other proposals~\cite{Hassler:2010a, Hassler:2011a}, the frequency shift associated with the fermion parity on the qubit islands enables us to close the loop 
	towards computational completeness for the topological qubits built of pairs of Majorana fermions.
	A logical qubit can be defined using two pairs of Majorana fermions, where the logical qubit states are $\ket{0}_{l} = \ket{00}$ and $\ket{1}_{l} = \ket{11}$. 
	Here, $\ket{0}$ and $\ket{1} = \gamma_{i} \ket{0}$ describe the two ground states of one of the Majorana fermion modes, 
	with the Majorana operators $\gamma_{i} = \gamma_{i}\hc$, see Ref.~\onlinecite{Alicea:2011a}.
	With this choice of logical qubit, braiding operations can be used to effect two-qubit gates as well as arbitrary $\pi/2$ single-qubit rotations~\cite{Alicea:2011a, Bonderson:2010a}. 
	In order to be able to perform arbitrary quantum gates, we need the additional capability to perform e.g., a $\pi/8$ phase gate on the logical qubit.
	Due to the energy difference between states of even and odd charge parity on the qubit islands, this can be realized by  
	simply moving one Majorana pair of the logical qubit to be manipulated onto one of the flux qubit islands for a time $t_{\text{Gate}}$, 
	as described in more detail in Refs.~\onlinecite{Hassler:2010a, Hassler:2011a}. 
	Then the state $\ket{1}_{l}$ will be different in energy from $\ket{0}_{l}$ by the frequency shift $\delta\omega_{r}$ induced in the resonator, and will then acquire a relative dynamical phase.
	For an energy separation between the two charge parity states of $25$~MHz, as is realized for the parameters used in Fig.~\ref{fig:ResFreq}, 
	a $\pi/8$-phase gate takes only $t_{\pi/8} = 2$~ns. In contrast to earlier proposals~\cite{Hassler:2010a, Hassler:2011a, Hyart:2013a}, 
	this gate when using a JCPM is protected to first order from charge as well as magnetic field fluctuations due to the operation of the JCPM at a triple sweet-spot. 
	
	The quality factor defined in Sec.~\ref{sec:Noise} allows us to give a bound on the fidelity of such a phase gate 
	effected by employing the energy splitting induced by the JCPM between different charge parity states.
	For simplicity, we assume that the only error stems from random frequency fluctuation of the resonator due to fluctuations in the external parameters of the JCPM.
	In this case, we use for the fidelity $ F = \abss{\braket{\psi}{\phi}} $, where $\ket\psi$ is the ideal state that we wanted to reach and $\ket\phi$ is the actual state reached during computation.
	We assume a $\theta$-phase gate, meaning an ideal final state of $\ket\psi = (\ket0 + \ee^{\ii \theta} \ket1)/\sqrt2$. 
	The phase of the actually realized state $\ket\phi$ deviates from $\theta$ by  $\delta\theta = \omega_{r} t_{\text{Gate}} / Q$ such that for small $\delta\theta$ we find
	\begin{align}
		F  \approx 1 - \frac14 \delta\theta^{2} \approx 1- 10^{-7} \,,
	\end{align} 
	where in the last step we assumed a $\pi/8$-phase gate and used the same parameters as above.
		
	It is interesting to also point out that the JCPM is not limited to use with Majorana fermions in superconductor-heterostructures. 
	Any operation that has a use for high fidelity QND charge-parity detection 
	would be a natural field of application~\cite{Beenakker:2004a, Trauzettel:2006a, Williams:2008a, Sun:2012a, Riste:2013a}. 
	As an example, the JCPM could be used as a quantum bus along the lines described in Ref.~\onlinecite{Bonderson:2011a} to entangle 
	topological Majorana qubits and charge qubits in semiconductor quantum dots or to transfer quantum information between the two.
	
	It can moreover be used as an ultra-sensitive, high-bandwidth dynamical charge detector. 
	Indeed, when biased in between the two charge sweet-spots at a region of maximum contrast and assuming standard homodyne reflection readout of the resonator, 
	we calculate a dynamical charge sensitivity in the range of $ 10^{-4} - 10^{-6} \text{e} / \sqrt{\text{Hz}}$
	with a bandwidth between $1 - 100$~MHz for the detection of fractional charges of $\delta q = 10^{-6}$~e. As explained in more details in~\ref{sec:Sensitivity},  conservative parameters and driving strengths have been used to obtain these figures of merit that are comparable to current state-of-the-art SET detectors~\cite{Aassime:2001a}.  

	In conclusion, we propose a new device based on superconducting circuits modulating the  frequency of a transmission-line cavity 
	as a function of an external charge. This device allows charge-parity detection and as such enables readout and manipulation of pairs of Majorana fermions 
	as topologically protected qubits. 
	Due to its operation at a triple sweet spot, with respect to two gate charges as well as applied magnetic flux, it is highly robust against environmental noise, 
	thus having the potential to preserve the protected character of Majorana fermion qubits for quantum computation.

	We thank N.~Didier, A.~Doherty and T.~Stace for discussions and acknowledge financial support by NSERC, CIFAR, and the Alfred P. Sloan Foundation.

\bibliography{Majorana_cQED}

\renewcommand\thefigure{\Alph{section}.\arabic{figure}} 
\renewcommand\theequation{\Alph{section}.\arabic{equation}}  
\renewcommand\thesection{Appendix \Alph{section}}
\setcounter{figure}{0}
\setcounter{equation}{0}
\setcounter{section}{0}

\section{Derivation of the JCPM Hamiltonian \label{sec:Lagrangian}}
	
\begin{figure}[tbp]
		\centering
		\includegraphics[width=0.8\columnwidth]{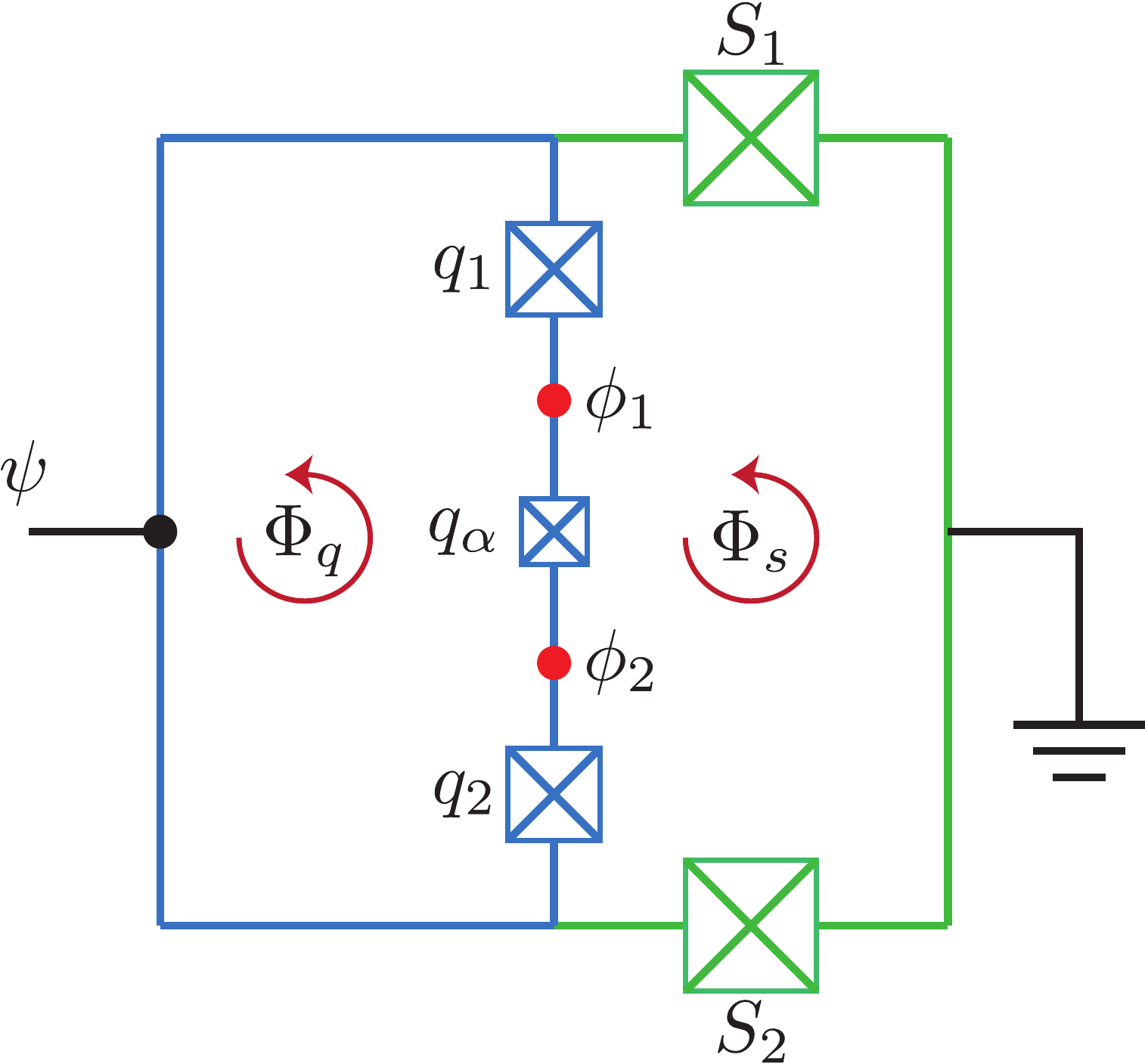}
		\caption{(Color online) Circuit diagram of the CMFQ device without the external gates. 
			Crosses indicate the position of Josephson junctions, which consists each of a nonlinear Josephson element as well as a capacitor in parallel.
			The labels next to the each of the elements correspond to the naming scheme used in the derivation of the Lagrangian. 
			The node fluxes at the red points will be the only remaining dynamical variables and correspond to the fluxes on the qubit islands. 
			For the full circuit we will connect voltage gates to these points, as in the inset in Fig.~\ref{fig:JPMFull}~(c).
			The black connection on the left leads to a distributed transmission line resonator and is at phase $\psi$ (whose dynamics are mainly determined by the resonator), 
			and the loops are threaded by the external fluxes $\Phi_{s}$ and $\Phi_{q}$ with the total external flux $\Phi_{\mathrm{x}} = \Phi_{s} + \Phi_{q}$.
		}
		\label{fig:JCM_Lagrangian}
	\end{figure}

	In this section we present the main steps in the derivation of the JCPM's Hamiltonian.
	The circuit diagram of the charge modulated inductance is shown in Fig.~\ref{fig:JCM_Lagrangian}. 
	To avoid unnecessary clutter, the external gates connecting to the islands are not represented in this figure. 
	As can be seen in Fig.~\ref{fig:JPMFull}~(c), these gates connect to the red dots in Fig.~\ref{fig:JCM_Lagrangian}.
	We assume equal SQUID junctions with $E_{J,S1} = E_{J,S2} = E_{J,s}$ and $C_{S1} = C_{S2} = C_{S}$. 
	Similarly, we take the qubit junction parameters as $E_{J,q1} = E_{J,q2} = E_{J,q}$ and $C_{q1} = C_{q2} = C_{q}$ 
	with the central `$\alpha$-junction' different by a factor $\alpha$: $E_{q_{\alpha}} = \alpha E_{q}$ and $C_{q_{\alpha}} = \alpha C_{q}$.
	We make these assumptions mainly for the sake of simplicity of notation, although it is somewhat important to the performance of the device as has already been discussed. 
		
	Following the approach outlined in Ref.~\onlinecite{Bourassa:2012a}, we find the total Lagrangian of the charge modulated flux qubit to be
	\begin{widetext}
		\begin{align}
			\mathcal L_{\text{CMFQ}} =& \tilde{C}_{S} \dot\psi^{2} 
					+ \frac12 \tilde{C}_{q} \Bigl[ \l( \dot\phi_{1} -\dot\psi \r)^{2} + \l( \dot\phi_{2} - \dot\psi \r)^{2} +\l( \dot\phi_{2} - \dot\phi_{1} \r)^{2} \Bigr]  \nn\\
				&+ E_{J,s} \Bigl[ \cos \psi  + \cos ( \psi + \Phi_{s} + \Phi_{q} ) \Bigr]
					+ E_{J,q} \Bigl[ \cos( \phi_{1} - \psi ) + \cos( \phi_{2} - \psi ) + \cos( \phi_{2} - \phi_{1} + \Phi_{q} ) \Bigr] \,.
		\end{align}
	\end{widetext}
	where, to simplify the notation in the main text, 
	we have defined the dimensionless phases $\psi = (2\pi / \Phi_{0} ) \: \tilde\psi$, $\phi_{1/2} = (2\pi / \Phi_{0} ) \: \tilde\phi_{1/2}$ and $\Phi_{s/q} = (2\pi / \Phi_{0} ) \: \tilde\Phi_{s/q}$
	and the renormalized capacitances $\tilde{C}_{s/q} = \l( \Phi_{0} /  2\pi \r)^{2} C_{s/q}$.	
	For simplicity of this calculation, 
	we now assume that the total external flux through both loops is a multiple of the flux quantum $\Phi_{0} = h/2e$, such that $\Phi_{\mathrm{x}} = \Phi_{s} + \Phi_{q} = 2\pi n$, 
	such that the SQUID is biased at a flux sweet spot. 
	With the substitution $\vphi_{\pm} = \phi_{1} \pm \phi_{2}$, we can then write the above Lagrangian as 
	\begin{widetext}
		\begin{align}
			\mathcal L_{\text{CMFQ}} =& \l( \tilde{C}_{S} + \tilde{C}_{q} \r) \dot\psi^{2} - \tilde{C}_{q} \dot\psi \dot \vphi_{+} 
					+ \frac14 \tilde{C}_{q} \l( \dot\vphi_{+}^{2} + \dot\vphi_{-}^{2} \r) + \frac12\alpha \tilde{C}_{q} \dot\vphi_{-}^{2} + 2 E_{J,s} \cos\psi \nn\\
				&+ 2 E_{J,q} \Bigl\{ \cos\psi \cos{\frac{\vphi_{+}}{2}}\cos{\frac{\vphi_{-}}{2}}  + \sin\psi \sin{\frac{\vphi_{+}}{2}}\cos{\frac{\vphi_{-}}{2}} \Bigr\} \nn\\
				&+ \alpha E_{J,q} \Bigl\{ \cos{\Phi_{q}} \cos{\vphi_{-}} + \sin{\Phi_{q}} \sin{\vphi_{-}} \Bigr\} \,.
			\label{eq:LJCPM2}
		\end{align}
	\end{widetext}
	Here and above, we have assumed that the arms of the SQUID have vanishing geometric inductance. Note that we have taken $\Phi_{\mathrm{x}} = \Phi_{s} + \Phi_{q} = 2\pi n$ to simplify the presentation of this particular expression. In general we however kept the dependence on the external flux.
	In the limit of infinite Josephson energy of the SQUID junctions, $E_{J,s} \rightarrow \infty$, the whole CMFQ circuit acts as a simple short to ground and we find $\psi = 0$.
	Here we assume large SQUID junctions $E_{J,s}$, such that $\psi\ll 1$, while still maintaining a significant inductive participation ratio $p_{L}$.
	As a result, the SQUID dynamics is frozen and its main effect on the resonator is to act as an effective nonlinear Josephson inductance of strength $2 E_{J,s}$.
	
	The full Lagrangian for the JCPM, consisting of a $\lambda / 4$ transmission-line resonator terminated by the CMFQ, 
	is then $\mathcal L_{\text{JCPM}} = \mathcal L_{\text{Res}_{0}} + \mathcal L_{\text{CMFQ}}$ with the uncoupled resonator part
	\begin{align}
		\mathcal L_{\text{Res}_{0}} = \int_{-\text{l}}^{0} dx \l( \frac{\Phi_{0}}{2\pi} \r)^{2} \l[ \frac{C_{0}}{2} \dot \psi(x)^{2} - \frac{1}{2 L_{0}} \l[ \partial_{x} \psi(x) \r]^{2} \r] \,,
	\end{align}
	where $C_{0}$, $L_{0}$ are capacitance and inductance per unit length and l, the length of the resonator. 
	Now we can identify the flux at the CMFQ-resonator port, indicated by the black dot in Fig.~\ref{fig:JCM_Lagrangian}, as given by the flux at one end of the resonator, $\psi = \psi(0)$.
	
	We can divide the full Lagrangian into parts describing the resonator, the charge-sensitive qubit and the interaction between the two as
	$\L_{\text{JCPM}} = \L_{\text{Res}} + \L_{\text Q} + \L_{\text{Q-Res}}$. Here, the Lagrangian describing the flux qubit (for simplicity again neglecting the additional gates on the two islands) is
	\begin{align}
		\L_{\text Q} =& \frac14 \tilde{C}_{q} \l( \dot\vphi_{+}^{2} + \dot\vphi_{-}^{2} \r) + \frac12\alpha \tilde{C}_{q} \dot\vphi_{-}^{2} 
				+ 2 E_{J,q}  \cos{\frac{\vphi_{+}}{2}}\cos{\frac{\vphi_{-}}{2}}  \nn\\
			&+ \alpha E_{J,q} \Bigl\{ \cos{\Phi_{q}} \cos{\vphi_{-}} + \sin{\Phi_{q}} \sin{\vphi_{-}} \Bigr\} \,,
		\label{eq:LQubit}
	\end{align}
	while for the renormalized resonator
	\begin{align}
		\L_{\text{Res}} = \L_{\text{Res}_{0}} + \l( \tilde{C}_{s} + \tilde{C}_{q} \r) \dot\psi^{2} + 2 E_{J,s} \cos\psi \,.
	\end{align}
	Most importantly, we find that the interaction takes the form
	\begin{align}
		\L_{\text{Q-Res}} &= - \tilde{C}_{q} \dot\psi \dot \vphi_{+} + 2 E_{J,q} \sin\psi \sin{\frac{\vphi_{+}}{2}}\cos{\frac{\vphi_{-}}{2}}  \nn\\
			+& 2 E_{J,q} \l( \cos\psi - 1 \r) \cos{\frac{\vphi_{+}}{2}}\cos{\frac{\vphi_{-}}{2}} \,.
		\label{eq:LJPM}
	\end{align}
	The terms proportional to $\sin\tilde\psi$ in this expression will lead to an Jaynes-Cummings-type interaction between the resonator and the qubit, 
	while the terms proportional to $( \cos\tilde\psi - 1)$ will renormalize the resonator parameters as well as introduce nonlinearities. 
	
	Focussing on the potential energy, we write the resonator phase variable as~\cite{Bourassa:2012a}
	\begin{align}
		\psi = \l( \frac{2\pi}{\Phi_{0}} \r) \sqrt{\frac{\hbar}{2 C_{r} \omega_{r,0} }} \l( a\hc + a \r) 
		\equiv \psi_{0} \l( a\hc + a \r) \,,
	\end{align}
	where $C_{r}$ is the total resonator capacitance and $\omega_{r,0} = 1/\sqrt{L_{r} C_{r}}$ the unloaded resonator frequency. 
	The operator $a$ is the standard annihilation operator for resonator photons.
	Taking the limit of large SQUID junctions, $E_{J,s} \gg 1$, the SQUID arms act on the resonator mainly as an inductive shunt to ground. 
	Then we find $\psi \ll 1$ and we can expand the expression Eq.~\eqref{eq:LJPM} around $\psi = 0$.
	To lowest order in $\psi$ we therefore have
	\begin{align}
		\L_{\text{Q-Res}} \approx & \beta_{1} \l( a\hc + a \r) \sum_{i,j} \bra i \op{\text{sc}} \ket j \ket j \bra i \nn\\
			+& \beta_{2} \l( a\hc + a \r)^{2} \sum_{i,j} \bra i \op{\text{cc}} \ket j \ket i \bra j \,,
		\label{eq:LJPM2}
	\end{align}
	where $\ket{i/j}$ are eigenstates of the flux qubit, $\beta_{1} = 2 E_{J,q} \psi_{0}$, $\beta_{2} = 2 E_{J,q} \psi_{0}^{2}$ and we have introduced the shorthand notation
	$\op{\text{sc}} = \sin{\frac{\vphi_{+}}{2}}\cos{\frac{\vphi_{-}}{2}}  $ and $\op{\text{cc}} = \cos{\frac{\vphi_{+}}{2}}\cos{\frac{\vphi_{-}}{2}} $.
	To determine which of the terms in Eq.~\eqref{eq:LJPM2} dominates, we calculate the matrix elements of the qubit coupling terms
	$\bra{i} \op{\text{sc}} \ket{j}$ and $\bra{i} \op{\text{cc}} \ket{j}$ for different qubit states. 
	Specifically, we are interested in how these terms change as a function of externally applied charges on the qubit islands.
	For this reason, we first have to determine how the presence of charges influence the flux qubit.
	
	From the Lagrangian $\L_{\text Q}$ we perform a Legendre transformation to arrive at the Hamiltonian $\mathcal H_{\text Q} = \sum_{i} q_{i} \dot \phi_{i} - \L_{\text Q}$. 
	Since the potential energy is left unchanged by this transformation, here we focus on the electrostatic (kinetic) energy terms and their dependence on gate charges on the two qubit islands. 
	First, in the absence of gate charges, we find
	\begin{align}
		\mathcal H_{\text{Kin}} = \frac{1}{ 2 C_{\Sigma}} \l\{ q_{1}^{2} (1+\alpha) + q_{2}^{2} (1+\alpha) + 2 \alpha q_{1} q_{2}\r\} \,,
		\label{eq:HKin}
	\end{align}
	with the island charges $q_{1} = \partial\L / \partial\dot \phi_{1}$ and $q_{2} = \partial\L / \partial\dot \phi_{2}$ and the total island capacitance $C_{\Sigma} = C_{q} (1+2\alpha)$.
	Modelling the effects of external charges on the qubit islands is then straightforward.  
	As illustrated in Fig.~\ref{fig:JPMFull}~(c) in the main text, at each of the two red points in Fig.~\ref{fig:JCM_Lagrangian} we add an additional capacitor $C_{G}$ 
	connected to a grounded voltage source on the other side.
	A simple calculation reveals that for each of these gates we have to supplement the Lagrangian with an additional  term of the form
	\begin{align}
		\L_{G_{i}} = \frac12 \tilde{C}_{G} \l( \phi_{i} - V_{G,i} \r)^{2} \,,
	\end{align} 
	with the island flux $\phi_{i}$ and the gate voltage $V_{G_{i}}$ applied to the gate capacitance of island $i$.
	With this addition, the kinetic energy terms in the Hamiltonian of Eq.~\eqref{eq:HKin} now take the modified form
	\begin{align}
		\mathcal H_{\text{Kin}} =&  \frac{(1+\alpha + \gamma)}{ 2 C'_{\Sigma}} \Bigl\{ (q_{1} + q_{G,1})^{2} + (q_{2} + q_{G,2})^{2} \Bigr\} \nn\\
			&+ \frac{\alpha}{C'_{\Sigma}} \l( q_{1} + q_{G,1} \r) \l( q_{2} + q_{G,2} \r) \,,
		\label{eq:HKinGate}
	\end{align}
	with the ratio between gate and qubit capacitances $\gamma = C_{G} / C_{q}$, the new total capacitance $C'_{\Sigma} = C_{q} (1+\gamma) (1+2\alpha+\gamma)$
	and where we have defined the charges $q_{G,i} = V_{G,i} C_{G}$.
	As usual, we have neglected terms $\propto q_{G,i}^{2}$ not contributing to the circuit dynamics.
	As expected, the additional gates cause an offset of the charge variables,	as well as a renormalization of the islands charging energy. 
	
	\begin{figure}[tbp]
		\begin{center}
			\includegraphics[width=0.9\columnwidth]{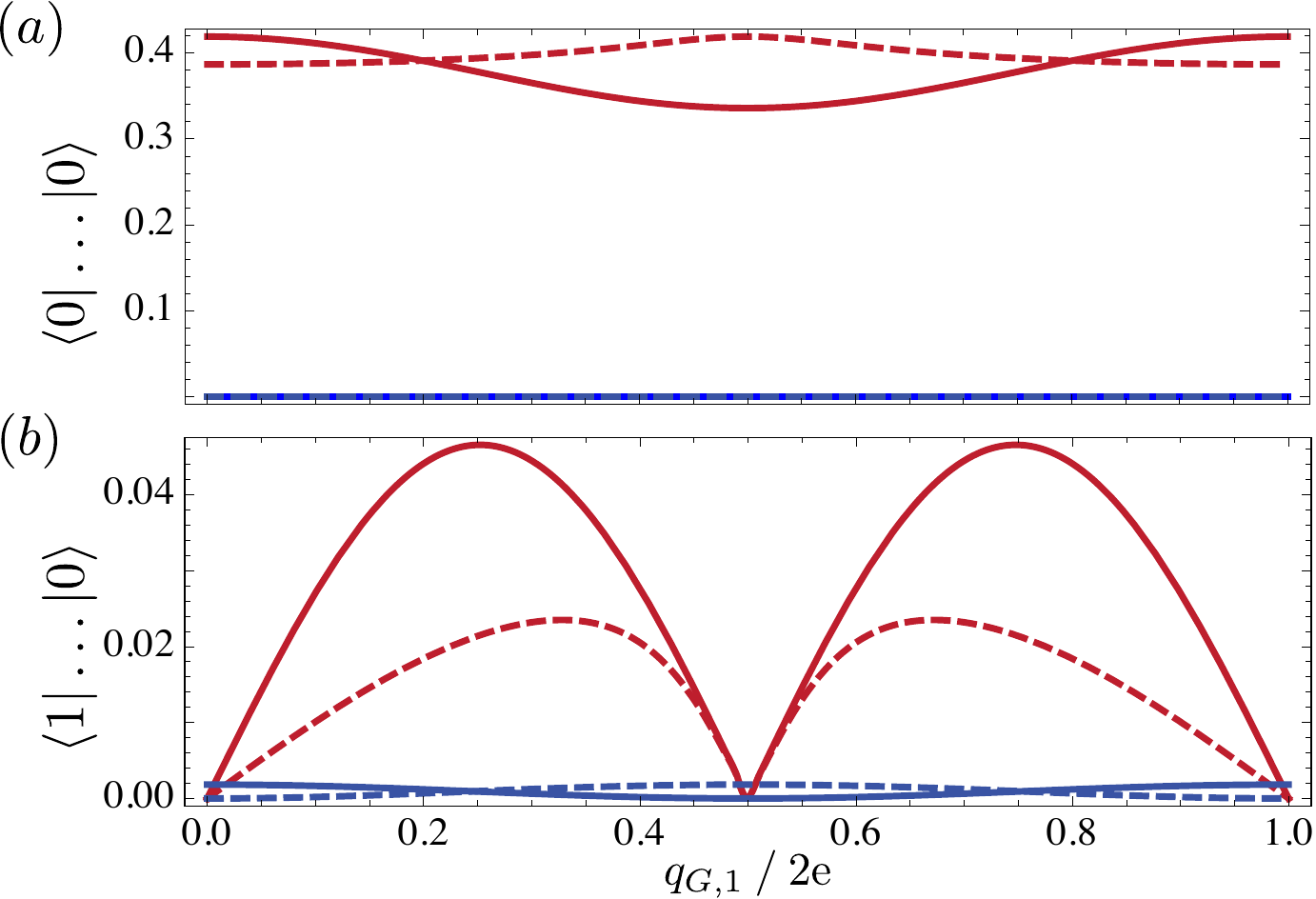}
			\caption{(Color online) Plot of the matrix elements of the CMFQ operators relevant for the interaction with the $\lambda/4$-resonator.
				\textbf{(a)} Even matrix elements $\bra0 \cos{\frac{\vphi_{+}}{2}}\cos{\frac{\vphi_{-}}{2}} \ket0 $ (red) 
				and $\bra0 \sin{\frac{\vphi_{+}}{2}}\cos{\frac{\vphi_{-}}{2}} \ket0$ (blue) 
				for two different charge bias points, $q_{G,2} = e$ (solid) and $q_{G,2} = 0$ (dashed).
				\textbf{(b)} Same for the uneven matrix elements $\bra1 \cos{\frac{\vphi_{+}}{2}}\cos{\frac{\vphi_{-}}{2}} \ket0 $ 
				and $\bra1 \sin{\frac{\vphi_{+}}{2}}\sin{\frac{\vphi_{-}}{2}} \ket0$. 
				Due to the symmetric circuit design the even terms, responsible for energy renormalisation, are two orders of magnitude larger than the uneven terms, 
				which would lead to an Jaynes-Cummings-like interaction between qubit and resonator degrees of freedom. 
			}
			\label{fig:MatrixElem}
		\end{center}
	\end{figure}

	To now see the influence on the resonator of additional charges on the qubit islands, we plot the matrix elements of the coupling terms in Eq.~\eqref{eq:LJPM2} for different qubit states.
	To do this, we calculate the eigenstates of the flux qubit numerically for a given set of parameters. 
	The qubit is here fully described by Eq.~\eqref{eq:LQubit} plus the additional gates to the islands from Eq.~\eqref{eq:HKinGate}.
	These eigenfunctions are then used to calculate and plot in Fig.~\ref{fig:MatrixElem} the relevant
	matrix elements of the two flux qubit operators $\op{\text{sc}}$ and $\op{\text{cc}}$ in Eq.~\eqref{eq:LJPM2}.
	For the chosen parameters (detailed below) and bias points at the sweet-spots, where $\Phi_{x} = \Phi_{s} + \Phi_{q} = 2\pi$ and $q_{G_{2}} = 0$ or e, 
	we find that the even matrix elements of the operator $\op{\text{sc}}$ are exactly zero and that $\bra i \op{\text{cc}} \ket j, \bra i \op{\text{sc}} \ket j \ll \bra i \op{\text{cc}} \ket i$.  
	Using the rotating wave approximation to neglect terms $\propto a^{2}$,
	we can then rewrite Eq.~\eqref{eq:LJPM2} as
	\begin{align}
		\L_{\text{Q-Res}} \approx& \l( a\hc + a \r) \sum_{i} g_{i} \l( \sigma_{+}^{(i)} + \text{h.c.} \r) + a\hc a \sum_{i} \delta\omega_{r,i} \ket i \bra i  \,,
	\end{align}
	where we have defined the Jaynes-Cummings coupling strengths $g_{i} \propto \bra{i+1} \op{\text{sc}} \ket{i} + \text{h.c.} $
	as well as the qubit state-dependent resonator pulls $\delta\omega_{r,i} \propto \bra i \op{\text{cc}} \ket i$. 
	The operator $\sigma_{+}^{(i)} = \ket{i+1}\bra i$ describes transition between adjacent qubit states.
	All of these parameters depend on the form of the flux qubit eigenstates and thus on the charge on the qubits islands.
	The coupling $g$ is only active nonperturbatively when the qubit-resonator detuning is small, $\Delta = \omega_{q} - \omega_{r} \sim g$.
	For our design we find $|g / \Delta| \ll 1$ as well as $\hbar \omega_{q} \ll k_{B} T$ (see main text) 
	and thus we expect the flux qubit to remain always in its ground state. 
	The main effect of the CMFQ on the resonator is then that of a charge tuneable inductance directly changing its resonance frequency by the amount $\delta\omega_{r,0}$.
	
	Finally, keeping higher order contributions in the Taylor expansion of $\cos\psi$ leads to Kerr type contributions $\sim K (a^\dag a)^2$ to the Hamiltonian. The magnitude of this nonlinear term is plotted in Fig.~\ref{fig:ResKerr}.

\section{Choice of JCPM parameters and additional results \label{sec:Parameters}}
	
	\setcounter{figure}{0}
	
	In order for the JCPM to work as a purely passive charge parity detector, several constraints on the circuit parameters have to be fulfilled.
	The basic idea behind the design is to couple a charge-sensitive flux qubit in such a way to a $\lambda/4$-resonator that its charge tuneability is only 
	reflected as an effective tuneable inductance. 
	In order for the qubit to remain a passive element, it has to remain in its ground state at all times. 
	Our symmetric coupling scheme, as shown in Fig.~\ref{fig:JPMFull}, suppresses direct exchange of energy between the resonator and the qubit. 
	Additionally we have to make sure that the qubit's level-splitting is always well above temperature, 
	$\hbar \omega_{q} \gg k_{B} T$, so that thermal excitations can be neglected. 
	As $\omega_q$ decreases exponentially with $\alpha$, and polynomially with the ratio of qubit Josephson to charging energy $E_{J,q} / E_C$~\cite{Chirolli:2006a}, 
	this restricts the values for these parameters in a realistic setting to $\alpha \approx 1$ and $E_{J,q} / E_C \sim 1 - 40$.
	Additionally, since the qubit energy is now a function of external charges, we have to make sure that this modulation does not 
	cause the qubit transition frequency to go below or even near $k_{B} T / h$ while still causing a large enough change in the resonator frequency.
	The qubit's sensitivity to charges increases with the inverse ratio $\l( E_{J,q} / E_{C} \r)^{-1}$, and thus a compromise between these two requirements has to be found. 
	Additionally the qubit Josephson energy $E_{J,q}$ cannot be very small as compared to the SQUID parameter $E_{J,s}$ so that the relative change in 
	effective inductance seen by the resonator is not too small.

	Throughout the main text and here, numerical simulations of the circuit response where performed with parameters well within the standard toolbox of superconducting circuit design.
	Specifically, for the resonator we assumed an unloaded fundamental frequency of $\omega_{r,0}/2\pi = 7.5$~GHz with the parameters
	\begin{align}
		C_{0} = 1.11 \times 10^{-10}~\text{F}/\text{m}\,, &\quad \quad L_{0}  = 2.78\times 10^{-7} \text{H}/\text{m}\,, \nn\\
		l &= 6~\text{mm}\,, \nn
	\end{align}
	while for the SQUID junctions we used 
	\begin{align}
		C_{J} = 5.17 \times 10^{-17}~\text{F}\,, \quad \quad E_{J,s}/2\pi = 350~\text{GHz}\,. \nn
	\end{align}
	Finally, the qubit was modelled with junctions of Josephson energy $E_{J,q}/2\pi = 200$~GHz while the charging energy of the islands was taken to be $E_{C}/2\pi = 20$~GHz.
	Neglecting the small gate capacitance, this translates into a qubit junction capacitance of $C_{q} = 3.19 \times 10^{-16}$~F.
	The qubit is also chosen to be symmetric, with all three junctions equal, $\alpha = 1$, and we assumed small gate capacitors of equal capacitance $C_{G} = 0.01 C_{q}$.
	The requirement of non-negligible charging energy $E_{C}$ of the qubit islands is the only parameter that is not realized routinely in todays experiments with flux qubits. 
	Specifically it requires that the coupling gate capacitances $C_{G}$ not be too large compared to $C_{q}$ such as not to lower $E_{C}$ too much.
	
	As mentioned above, we want to bias the SQUID at a flux sweet spot such that its dynamics is frozen and it acts only as an effective Josephson inductance. 
	To achieve this, the total external flux through the CMFQ loop must be an integer multiple of the flux quantum $\Phi_{0}$, resulting in $\Phi_{x} = 2n \pi$ with integer $n$.
	Additionally, for best results, we want to bias the flux qubit at one of its flux sweet spot, $\Phi_{q} = n \pi$. 
	
	\begin{figure}[tbp]
		\centering
		\includegraphics[width=0.9\columnwidth]{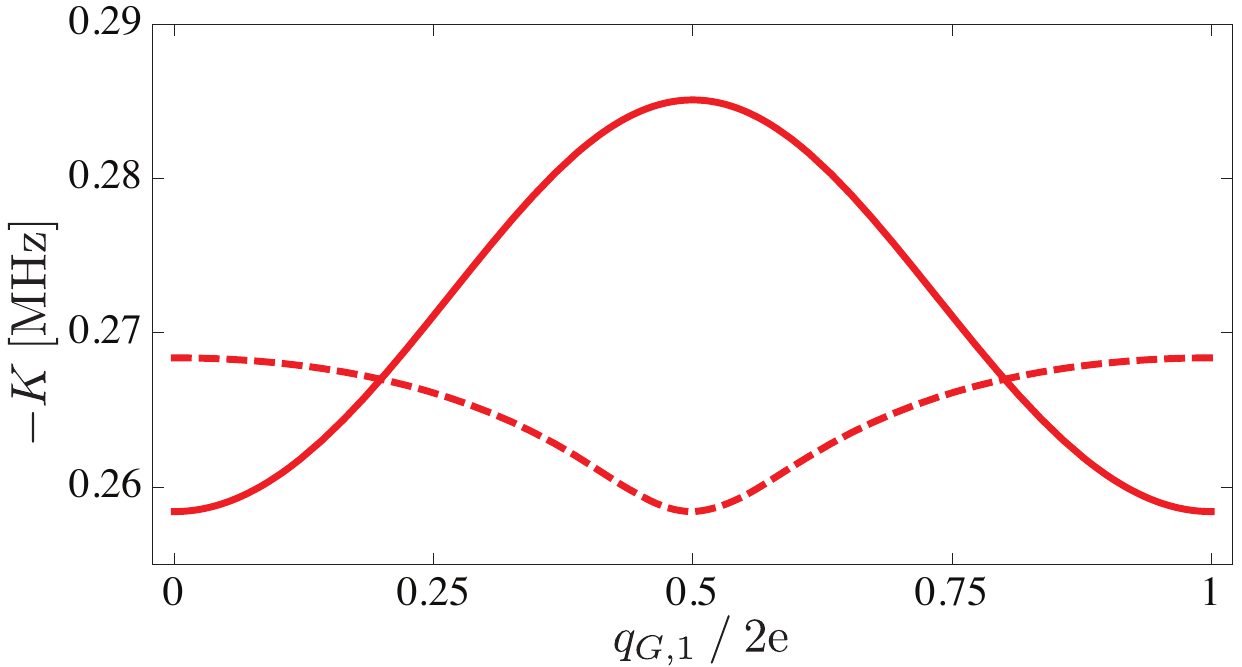}
		\caption{(Color online) Kerr nonlinearity $K$ as a function of gate charge $q_{G,1}$ at the flux sweet spot of the device, $\Phi_{q} = \Phi_{0} / 2$.
			The solid line is with charge bias on the second qubit island, $q_{G,2} = $e, the dashed line for no charge bias, $q_{G,2} = 0$
			With the chosen parameters, the Kerr nonlinearity is modulated by $\sim 10\%$ and is of the right order to enable enhanced readout schemes~\cite{Mallet:2009a, Ong:2011a}.
		}
		\label{fig:ResKerr}
	\end{figure}
	
	Finally, Fig.~\ref{fig:ResKerr} shows the induced resonator nonlinearity $K$ as a function of gate charge $q_{G,1}$ for two charge bias points, 
	$q_{G,2} = 0$ and $q_{G,2} = $~e, on the second qubit island.
	In both cases, the nonlinearity $K$ is of the order of $\sim -200$~kHz.
	This is the same order of magnitude than the nonlinearities used in single-shot bifurcation readout~\cite{Mallet:2009a} and qubit measurement with nonlinear resonators~\cite{Ong:2011a}.
	In the same way, the nonlinearity introduced by the CMFQ could thus be used to improve the fidelity of charge parity measurements.

\section{Aharonov-Casher effect in flux qubits \label{sec:AharonovCasher}}
	
	\setcounter{figure}{0}
	
	\begin{figure}[tbp]
		\begin{center}
			\includegraphics[width=.9\columnwidth]{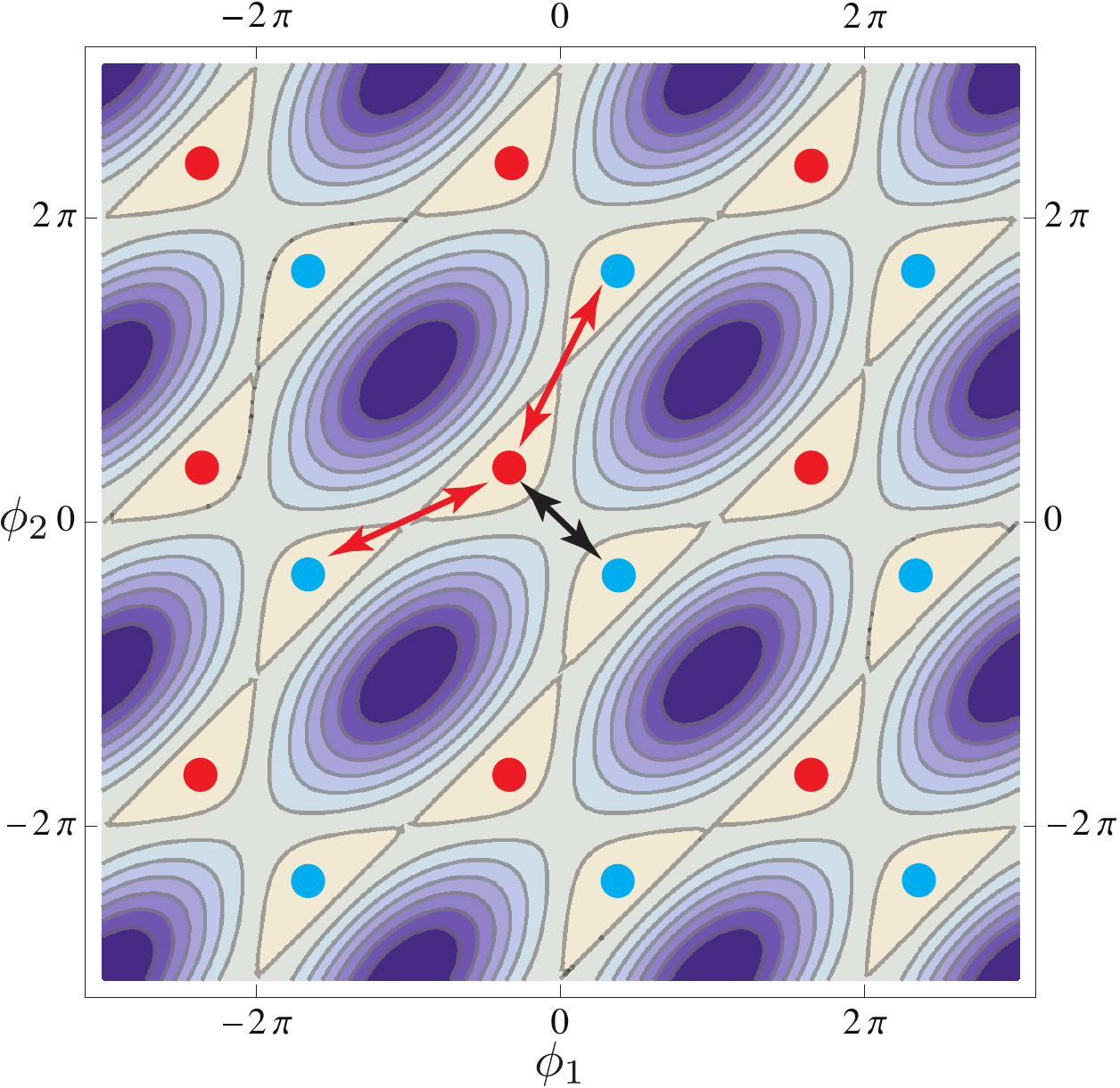}
			\caption{(Color online) Potential of a charge tuneable flux qubit with $\alpha = 1.2$ as function of the islands fluxes $\phi_{1/2}$ 
				and at its flux sweet spot, $\Phi_{q} = \pi$.
				Points of the same color denote potential minima corresponding either clockwise or counter-clockwise current through the qubit loop.
				All points of the same color are equivalent since they only differ by multiples of $2\pi$ in one of the phases. 
				Arrows indicate tunneling paths leading to hybridization of current states. For further details, see text.
			}
			\label{fig:ACEffect}
		\end{center}
	\end{figure}
	
	Here we give a brief explanation of the charge-tuneability of the flux qubit transition (and thus of the effective inductance it provides to the circuit) 
	in terms of the Aharonov-Casher effect. More details can be found in Refs.~\onlinecite{Chirolli:2006a, Tiwari:2007a}.
	Fig.~\ref{fig:ACEffect} shows the potential landscape for a charge tuneable flux qubit according to Eq.~\eqref{eq:LQubit}.
	Colored dots indicate potential minima whose groundstates have either clockwise or counter-clockwise current in the qubit loop. 
	Tunneling between potential minima leads to hybridization of current states and, close the flux sweet spot, is the main contribution to the qubit energy splitting.
	For a standard flux qubit with $\alpha < 1$, the lowest barrier between two minima is indicated by the black arrow in Fig.~\ref{fig:ACEffect}.
	Tunneling along this path corresponds to a change in the variable $\vphi_{-} = \phi_{1} - \phi_{2}$. 
	On the other hand, for a charge tuneable flux qubit, where $\alpha \gtrsim 1$, which is the situation depicted in the figure, the two barriers in the direction $\vphi_{+} = \phi_{1} + \phi_{2}$,
	indicated by the red arrows, are lower in energy. 
	There then exist two competing tunneling paths that are contributing to the qubit level splitting.
	If charges are present on one of the qubit islands, the different paths acquire different phases due to the AC effect. 
	Constructive of destructive interference of the two paths is then possible depending on the number of charges involved. 
	Thus the tunneling amplitude and therefore the qubit level splitting can be tuned by application of charges to the qubit islands.
	
\section{Dynamical Charge Sensitivity \label{sec:Sensitivity}}
	
	\setcounter{figure}{0}
	
	\begin{figure}[tbp]
		\begin{center}
			\includegraphics[width=.9\columnwidth]{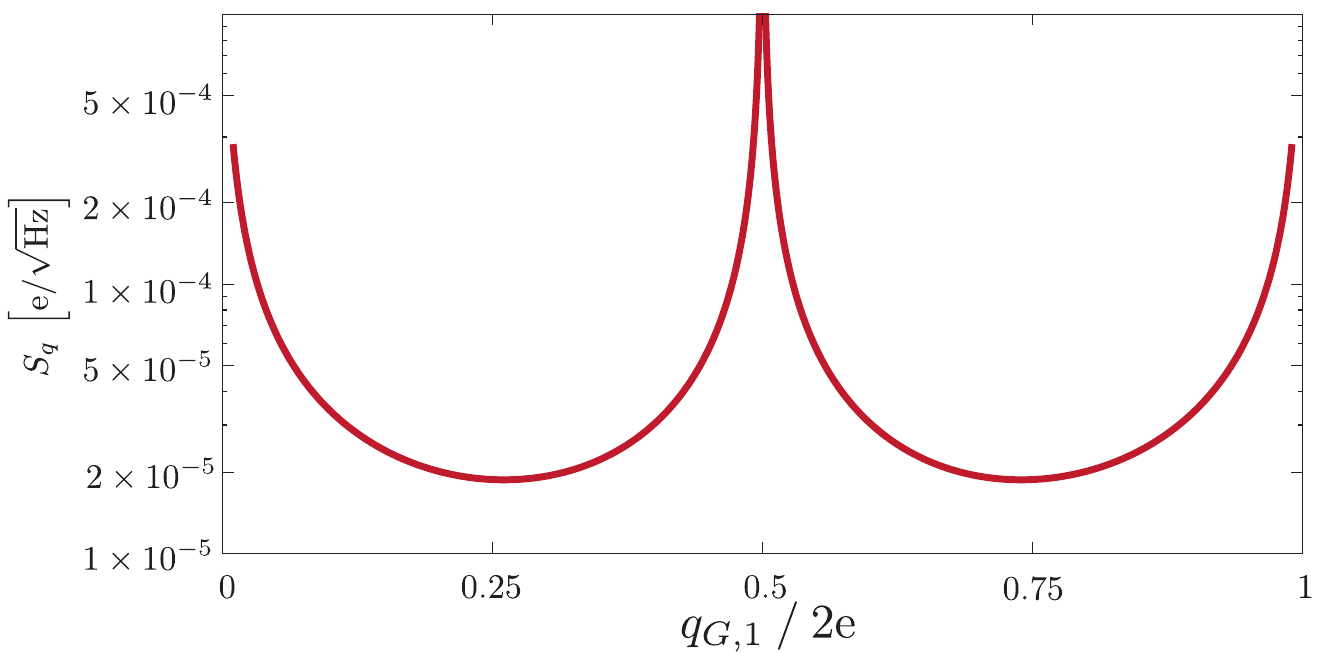}
			\caption{(Color online) Dynamical charge sensitivity $S_{q}$ as a function of bias charge $q_{G,1}$ on the lower flux qubit island 
				with the upper island biased at the operating point for charge parity detection, $q_{G,2} = e$, $\Phi_{x} = 2\pi$.
				We take the resonator photon loss rate $\kappa = 10$~MHz and an intermediate number of measurement photons $n=10$.
			}
			\label{fig:Sq}
		\end{center}
	\end{figure}

	When biased away from its charge sweet spots, the JCPM can be used as a highly sensitive dynamical charge detector.
	As a figure of merit, we define a charge sensitivity 
	assuming a standard homodyne measurement~\cite{QuantumOptics} of the reflected phase of a microwave signal to determine the resonance frequency of the JCPM.
	Following Ref.~\onlinecite{Blais:2004a}, we write the sensitivity of such a measurement as 
	\begin{align}
		S_{\vphi} = \delta \vphi \sqrt{T_{m}} = \sqrt{ \frac{1}{\kappa n} } \,,
		\label{eq:SvPhi}
	\end{align}
	where $\delta \vphi$ is the variation in phase that can be resolved in the measurement time $T_{m}$. 
	Assuming a reflection measurement with a single port, i.e. no transmission losses of photons in the cavity, we find $T_{m} = 1 / (\kappa n \delta\vphi^{2})$ 
	and arrive at the second equality in Eq.~\eqref{eq:SvPhi}.
	The sensitivity depends on the resonator leakage rate $\kappa$ at which photons arrive at the measurement apparatus, 
	and the number of photons $n$ determined by the measurement driving strength.
	
	For small variations in resonator frequency $\delta\omega_{r}$, the change in reflected phase is linear as $\delta\vphi \approx \delta\omega_{r} / \kappa$.
	In the JCPM, the resonance frequency depends directly on the external charge and we can write $\delta\omega_{r} = \dvpart{\omega_{r}}{q} \delta q$, 
	where $\delta q$ is a change in applied gate charge, which we want to measure. 
	We thus write
	\begin{align}
		S_{\vphi} = \frac{1}{\kappa} \dvpart{\omega_{r}}{q} \delta q \sqrt{T_{m}} = \frac{1}{\kappa} \dvpart{\omega_{r}}{q} S_{q} \,,
	\end{align}
	where we identified the dynamical charge sensitivity $S_{q} = \delta q \sqrt{T_{m}}$.
	Hence, we arrive at the equation 
	\begin{align}
		S_{q} = \sqrt{\frac{\kappa}{n}} \l( \dvpart{\omega_{r}}{q} \r)^{-1} \,,
	\end{align}
	defining the dynamical sensitivity of a charge measurement using a JCPM with measurement of the reflected phase of the driving signal.
 	In Fig.~\ref{fig:Sq} we plot this sensitivity as a function of charge bias point on the first qubit gate,  
	for the same operating point as above and assuming standard circuit QED measurement parameters of $\kappa / 2\pi = 10$~MHz and a moderately strong drive 
	with a total number of resonator photons $n=10$. The minimum sensitivity is reached for biasing in between the two sweet spots at a gate charge of 
	$q_{G,1} = 0.5$~e and is of the order of $2\times 10^{-5}$~$e/\sqrt{\text{Hz}}$ for these parameters.
	Obviously, stronger driving of the resonator will allow for faster measurement and thus increased sensitivity. 
	Varying the number of resonator photons $n$ between $1$ and $100$, the sensitivity ranges between $S_{q} = 10^{-4} - 10^{-6}$~e$/\sqrt{\text{Hz}}$.
	
	To give a bandwidth of charge detection we take a closer look at the measurement time $T_{m}$.
	We write again $\delta\omega_{r} = \dvpart{\omega_{r}}{q} \delta q$ and thus find for the detection bandwidth, i.e. the inverse measurement time
	\begin{align}
		\text{BW} = \frac{1}{T_{m}} = \frac{n}{\kappa} \l( \dvpart{\omega_{r}}{q} \r)^{2} \delta q^{2} \,.
	\end{align}
	With the same parameters as above we find the bandwidth for detection of fractional charges $\delta q = 10^{-6}$~e as $BW = 1- 100$~MHz 
	when varying the resonator drive power between $n = 1 - 100$.
 	
	Note that we have so far assumed perfect measurement efficiency, $\eta = 1$, which is obviously not fulfilled in experiments. 
	However, the only change from the above equations will be to raise the sensitivity as $S_{q} \sim 1 / \sqrt\eta$ and reduce the bandwidth as BW$ \sim \eta$.	

\end{document}